\documentstyle[a4,12pt,amssymb,amsthm,latexsym,amsfonts]{article}

\newcommand{\Med}[1]{\left\langle #1 \right\rangle}
\newcommand{\med}[1]{\langle #1 \rangle} 
\newtheorem{theorem}{Theorem}

\title{
Quadratic replica coupling in the Sherrington-Kirkpatrick mean field
spin glass model}
 
\author{Francesco Guerra
\footnote{\ 
e-mail: {\tt francesco.guerra@roma1.infn.it}}\\{\small {\itshape Dipartimento di Fisica, Universit\`a di Roma ``La
Sapienza''}}\\ {\small {\itshape Istituto Nazionale di Fisica Nucleare, Sezione di
Roma1}}\\{\small {\itshape Piazzale Aldo Moro, 2, I-00185 Roma, Italy}}\\
\and
Fabio Lucio Toninelli
\footnote{\ 
e-mail: {\tt toninell@cibs.sns.it}}
\\{\small {\itshape Scuola Normale Superiore,
Piazza dei Cavalieri 7, 56126 Pisa, Italy}}\\
{\small {\itshape Istituto Nazionale di Fisica Nucleare, Sezione di Pisa}}}
 \date{\today}
\begin{document}

\maketitle


\begin{abstract}
\noindent
We develop a very simple method to study the high temperature, or equivalently
high external field, behavior of the
Sher\-ring\-ton-Kirkpatrick mean field spin glass model. The basic idea is to
couple two different replicas with a quadratic term, trying to push out the two
replica overlap from its replica symmetric value. In the case of zero external
field, our results reproduce the well known validity of the annealed
approximation, up to the known critical value for the temperature. In the case
of nontrivial external field, we can prove the validity of 
the Sher\-ring\-ton-Kirkpatrick replica symmetric solution up to a line, which
falls short of the Almeida-Thouless line, associated to the onset of the
spontaneous replica symmetry breaking, in the Parisi Ansatz. The main
difference with the method, recently developed by Michel Talagrand, is that we
employ a quadratic coupling, and not a linear one. The resulting flow
equations, with respect to the parameters of the model, turn out to be much
simpler, and more tractable. As a straightforward application of cavity
methods, we show also how to determine free energy and overlap fluctuations, in
the region where replica symmetry has been shown to hold. It is a major open
problem to give a rigorous mathematical treatment of the transition to replica symmetry
breaking, necessarily present in the model.
\end{abstract}

\section{Introduction}

The mean field spin glass model, introduced by
Sherrington and Kirkpatrick in \cite{SK}, is here considered in the high
temperature regime, or, equivalently, for a large external field. It is very
well known, on physical grounds, that in this region the replica symmetric
solution holds, as shown for example in \cite{MPV}, and references quoted
there. However, due to the very large fluctuations present in the model, it
is not so simple to give a complete, mathematically
rigorous, characterization  of this region, especially when there are 
external fields. Rigorous work on this subject include \cite{ALR}, \cite{FZ},
\cite{locarno}, \cite{S}, \cite{CN}, \cite{MT1}. 
For other rigorous results on the structure of the model,
we refer to \cite{PS}, \cite{MT}, \cite{FG}, \cite{AC}, \cite{GG}, \cite{BR}, 
\cite{SUM}.

The method developed in \cite{MT1}, by Michel Talagrand, is particularly
interesting. The starting point is given by the very sound physical idea
that the spontaneous replica symmetry breaking phenomenon can be understood
by exploring the properties of the model, under the application of
auxiliary interactions, which explicitly break the replica symmetry. In
\cite{MT1}, the replica symmetric solution is shown to hold  in a region,
which ({\it probably}) coincides with the region found in the theoretical
physics literature, as shown for example in \cite{MPV}, {\sl i.e.} up to the 
Almeida-Thouless line \cite{AT}.

The main tool in Talagrand's treatment is an additional minimal replica
coupling, linear in the overlap between two replicas. Then, a kind
of quadratic stability for the so modified free energy leads immediately, through a generalization of the methods developed in \cite{SUM}, to
establish the validity of the replica symmetric solution in a suitable parameter region.

Here we propose a very different strategy, by introducing a quadratic
replica coupling, attempting to push the overlap away from its replica
symmetric value. In a sense, our method is the natural extension, with
applications, of the ideas put forward in \cite{SUM}, where sum rules were
introduced for the free energy, by expressing its deviation from the replica
symmetric solution in terms of appropriate quadratic fluctuations for the
overlap. We choose exactly these quadratic fluctuation terms to act as
additional interaction between two replicas, thus explicitly breaking replica
symmetry. Then, a generalization of the sum rules, given in \cite{SUM}, for
this modified model, allows us immediately to prove that the free energy of
the original model converges, in the infinite volume limit, to its replica
symmetric value, at least in a parameter region, explicitly determined.

The organization of the paper is as follows. In Section 2, we recall the
basic definitions of the mean field spin glass model, and introduce the
overlap distribution structure. As a first introduction of our method of
quadratic coupling, in Section 3, we treat the well known case of zero
external field, by showing that the annealed approximation holds, in the
infinite volume limit, up to the true critical inverse temperature
$\beta_c=1$. Our proof shows explicitly that there is a strong connection
between the critical value of the transition temperature for the zero
external field model, and the analogous, and numerically
equivalent, temperature for the well known ferromagnetic Curie-Weiss mean
field model.

In Section 4, we consider the model with external field, and introduce the
associated model with quadratic replica coupling. Then, simple stability
estimates give immediately the convergence of the free energy to its replica
symmetric value, in a suitable, well defined, region of the parameters. 

Section 5 reports about results on the free energy and overlap
fluctuations, in our determined replica symmetric region. We also sketch the
method of proof, based on cavity considerations, as developed for example
in \cite{MPV} and \cite{locarno}. A more complete treatment will be found in a
forthcoming paper \cite{GT}.

Finally, Section 6 is dedicated to a short outlook about open problems and further
developments.

For the relevance of the mean field spin glass model for the understanding of the physical properties of realistic spin glasses, we refer to \cite{MPR}, but see also 
\cite{NS}.

\section{The general structure of the mean field spin glass model}

The generic configuration of the mean field spin glass model is defined by Ising spin variables
$\sigma_{i}=\pm 1$,  attached to each site $i=1,2,\dots,N$. The external quenched disorder is
given by the $N(N-1)/2$ independent and identical distributed random
variables $J_{ij}$, defined for each couple of sites. For the sake of simplicity, we assume each $J_{ij}$ to be a centered
unit Gaussian with averages
$$E(J_{ij})=0,\quad E(J_{ij}^2)=1.$$
The Hamiltonian of the model, in some external field of strength $h$,  is given by
\begin{equation}\label{H}
H_N(\sigma,J)=-{1\over\sqrt{N}}\sum_{(i,j)}J_{ij}\sigma_i\sigma_j
-h\sum_{i}\sigma_i.
\end{equation}
The first sum extends to all site couples, an the second to all sites. The
normalizing factor ${1/\sqrt{N}}$ is typical of the mean field character of
the model, and guarantees a good thermodynamic 
limit for the free energy per spin, i.e., the existence of a finite and
non trivial limit for the free energy as $N\to\infty$. 
The first term in (\ref{H}) is a long range random two body interaction,
while the second represents the interaction of the spins with a fixed 
external magnetic field $h$.

For a given inverse temperature $\beta$, we introduce the disorder dependent
partition function $Z_{N}(\beta,J)$, the (quenched average of the) free energy per site
$f_{N}(\beta)$, the internal energy per site
$u_{N}(\beta)$, the Boltzmann state $\omega_J$, and the auxiliary function $\alpha_N(\beta)$,  
according to the definitions
\begin{equation}\label{Z}
Z_N(\beta,J)=\sum_{\sigma_1\dots\sigma_N}\exp(-\beta
H_N(\sigma,J)),
\end{equation}
\begin{equation}\label{f}
-\beta f_N(\beta)=N^{-1} E\log Z_N(\beta,J)=\alpha_N(\beta),
\end{equation}
\begin{equation}\label{state}
\omega_{J}(A)=Z_N(\beta,J)^{-1}\sum_{\sigma_1\dots\sigma_N}A\exp(-\beta
H_N(\sigma,J)), 
\end{equation}
\begin{equation}\label{u}
u_N(\beta)=N^{-1}E\omega_{J}(H_N(\sigma,J))=
\partial_{\beta}\bigl(\beta
f_N(\beta)\bigr)=-\partial_{\beta}\alpha_N(\beta), 
\end{equation}
where $A$ is a generic function of the $\sigma$'s. In the
notation $\omega_J$, we have stressed the dependence of the Boltzmann
state on the external noise $J$, but, of course, there is also
a dependence on $\beta$, $h$ and $N$.

We are interested in the thermodynamic limit $N\to\infty$.

Let us now introduce the important concept of replicas. Consider a generic number $s$ of independent copies
of the system, characterized by the Boltzmann
variables $\sigma^{(1)}_i$, $\sigma^{(2)}_i$, $\dots$,
distributed according to the product state
$$\Omega_J=\omega^{(1)}_J \omega^{(2)}_J \dots\omega^{(s)}_J,$$
where all $\omega^{(\alpha)}_J$ act on each one
$\sigma^{(\alpha)}_i$'s, and are subject to the {\sl
same} sample $J$ of the external noise. Clearly, the {\sl Boltzmannfaktor} for the replicated system is given by
\begin{equation}
\exp\left(-\beta
(H_N(\sigma^{(1)},J)+H_N(\sigma^{(2)},J)+\dots+H_N(\sigma^{(s)},J))\right).
\end{equation}

The overlaps between two replicas $a,b$ are
defined according to
$$q_{ab}(\sigma^{(a)},\sigma^{(b)})={1\over
N}\sum_{i}\sigma^{(a)}_i\sigma^{(b)}_i,$$
and they satisfy the obvious bounds
$$-1\le q_{ab}\le 1.$$

For a generic smooth function $F$ of the overlaps, we
define the $\langle\rangle$ averages
$$\langle F(q_{12},q_{13},\dots)\rangle=E\Omega_J\bigl(F(q_{12},q_{13},\dots)\bigr),$$
where the Boltzmann averages $\Omega_J$ acts on the replicated $\sigma$
variables, and $E$ is the average with respect to the external noise $J$. 

We
remark here that the noise average $E$ introduces correlations between different
groups of replicas, which would be otherwise independent under the Boltzmann
averages $\Omega_J$, as for example $q_{12}$ and $q_{34}$.  

The $\langle\rangle$ 
averages are obviously invariant under permutations of the replicas.

Overlap distributions play a very important role in the theory. For example, a simple direct calculation shows \cite{MPV}, \cite{FG} that
\begin{equation}\label{der}
\partial_{\beta}\alpha_N(\beta)={\beta\over2}\bigl(1-\langle q_{12}^{2}\rangle\bigr).
\end{equation} 

In order to introduce our treatment, based on flow equations with respect to the parameters of the theory, it is convenient to start from a {\sl Boltzmannfaktor}
given by 

\begin{equation}\label{Bf}
\exp\,\left(\sqrt{\frac tN}\sum_{(i,j)}J_{ij}\sigma_i\sigma_j
+\beta h\sum_{i}\sigma_i +\sqrt{x}\sum_{i}J_{i}\sigma_i\right).
\end{equation}
Notice that we have introduced an auxiliary additional one body random interaction
ruled by the strength $\sqrt{x}$, $x\ge0$, and $N$ quenched independent and
identically distributed centered unit Gaussian random variables $J_i$, so that
%
$$E(J_{i})=0,\quad E(J_{i}^2)=1.$$
%
In order to get the original model, we have to put $x=0$ at the end.
Moreover, we have written $\beta=\sqrt{t}$, with $t\ge0$. The variables $t$, and
$x$, will play the role of time variable, and space variable, respectively, in our flow equations. Now we define the partition function $Z$ by using the {\sl Boltzmannfaktor} (\ref{Bf}), and the auxiliary function $\alpha_{N}(x,t)$ in the form
%
$$\alpha_{N}(x,t)=N^{-1} E\log Z_N.$$
%
Then, as in the proof of (\ref{der}),  we have
\begin{eqnarray}
&&\partial_{t}\alpha_{N}(x,t)={1\over4}\bigl(1-\langle q_{12}^{2}\rangle\bigr)\\\label{dder}
&&\partial_{x}\alpha_{N}(x,t)=
{1\over 2}\bigl(1-\langle q_{12}\rangle\bigr). 
\end{eqnarray}
It is very simple to calculate explicitly the average $N^{-1}E\log Z_N$ for $t=0$, at a generic strength $x_0$ of the one
body random interaction. In fact, at $t=0$, the
interaction factorizes, and the spins at different sites
become independent. Therefore we have
\begin{equation}\label{t0}
\alpha(x_0,0)=\log 2 + \int
\log \cosh (\beta h + z \sqrt x_0) d\mu(z),
\end{equation}
independently of $N$, where $d\mu$ is the centered unit
Gaussian, representing each single $J_i$,
$$d\mu(z)=\exp(-\frac{z^2}2) dz/\sqrt{2\pi}.$$
Starting from (\ref{t0}),  (\ref{dder}) at $t=0$, we can  immediately  calculate the order parameter  $\overline{q}(x_0)$ according to 
%
$$\overline{q}(x_0)=\langle q_{12}\rangle (x_0,0)=\int\tanh^2(\beta
h + z \sqrt{x_0}) d\mu(z).$$

Let us now consider consider  linear trajectories at constant
velocity given by
\begin{equation}\label{traj0}
x(t)=x_0 - {\overline q}(x_0)t,
\end{equation}
where $x_0$ is the initial starting point.
It is immediate to verify that there is only one
trajectory passing for a generic point $(x,t)$, $x\ge0$,
$t\ge0$, {\sl i.e.} the initial point is uniquely determined by
$(x,t)$, and so is for $\overline q$. In fact, the following
Theorem holds.
\begin{theorem}
\label{unique}
Consider first the case of non zero external field $h$. Then, for a generic point $(x,t)$, $x\ge0$, $t\ge0$, there
exists a unique $x_0(x,t)$ such that
%
$$x=x_0(x,t)-{\overline q}(x_0(x,t))t,$$
%
and a unique ${\overline q}(x,t)={\overline q}(x_0(x,t))$, such that
%
$${\overline q}(x,t)=\int\tanh^2\bigl(\beta
h + z \sqrt {x+{\overline q}(x,t)t}\bigr) d\mu(z).$$
%
If $h=0$, then invertibility is assured, with some $x_0(x,t)>0$, in the region $x\ge0$, $t\ge0$, with the exclusion of the segment $x=0$, $0\leq t\leq 1$.  
\end{theorem}
The proof is very simple, and can be found in \cite{SUM}.

By following the methods of  \cite{SUM}, sum rules connecting $\alpha$ and its  SK approximation can be easily found by using transport equations. They involve overlap fluctuations. In fact, let us consider $\alpha$ along the trajectories in (\ref{traj0}), {\sl i.e.} $\alpha(x(t),t)$. An easy calculation gives
\begin{equation}\label{veri}
\frac {d}{dt}\alpha_N(x(t),t)=\frac{1}{4}\bigl(1-\overline{q}(x_0)\bigr)^2-\frac{1}{4} \langle(q_{12}-\overline{q})^2\rangle.
\end{equation}
Now we integrate along $t$, take into account the initial condition (\ref{t0}), and the invertibility of (\ref{traj0}), and find the sum rule
\begin{equation}\label{sum}
{\overline\alpha}(x,t)=\alpha_{N}(x,t)+\frac{1}{4}\int_0^t \langle(q_{12}-\overline{q})^2\rangle_{x(t'),\,t'} dt'. 
\end{equation}
Here, we have defined  the replica
symmetric Sherrington-Kirkpatrick solution \cite{SK}, \cite{MPV}, in the form
\begin{equation}\label{aSK}
{\overline\alpha}(x,t)=\log 2 + \int
\log \cosh \bigl(\beta h + z \sqrt x_0\bigr) d\mu(z) + \frac{t}{4}\bigl(1-\overline{q}(x_0)\bigr)^2.
\end{equation}
A very simple, but important, consequence of the sum
rule is that $\alpha_N$ is dominated by its replica
symmetric solution, uniformly in $N$,
\begin{equation}\label{dom}
\alpha_N(x,t)=N^{-1} E \log Z_N \le {\overline\alpha}(x,t).
\end{equation}
This is a simple consequence of the positivity of the
term under integration in (\ref{sum}).

Now, we are ready to explain our method of quadratic coupling, starting with the simple case of zero external field, and then going to the case of nontrivial external fields.

\section{Quadratic coupling for zero external field}

The high temperature region ($\beta<1$) of the 
zero external field SK model is a very particular case where everything can 
be computed. As it is well known \cite{ALR}, in this case the annealed approximation is exact in the infinite volume limit. In fact, we have
\begin{eqnarray}
\label{ann}
\lim_{N\to\infty}\frac1N E\ln Z_N(t,J)={\overline \alpha}(t)\equiv
\ln2+\frac t4=\frac1N\ln E Z_N(t,J)+\frac t{4N}.
\end{eqnarray}

In this section we give a new proof of Eq. (\ref{ann}), based on  
sum rules for the free energy.  Our method is very simple and can be easily 
extended to the case of nontrivial external field,  considered in the next
Section.

When $h=0$, $x=0$, then also  $\bar q=0$ and $x_0=0$. Then, the sum rule 
in (\ref{sum}) reads
\begin{equation}
\label{rs}
\alpha_N(t)={\overline\alpha}(t)-\frac14\int_0^t\med{q_{12}^2}_{t'}\,dt'.
\end{equation}
The presence of $\med{q_{12}^2}$, as order parameter, suggests to couple
two replicas with a term proportional to the square of the overlap, the 
corresponding partition function being
\begin{equation}\label{tildez}
\tilde Z_N(t,\lambda,J)=\sum_{\{\sigma,\sigma'\}}
\exp\,\left(\sqrt{\frac tN}\sum_{(i,j)}J_{ij}(\sigma_i\sigma_j+
\sigma'_i\sigma'_j)+\frac\lambda2 N q_{12}^2\right),
\end{equation}
with $\lambda\geq0$.
The effect of the added term is to give a larger weight to the 
configurations having $q_{12}\neq0$, thus favoring nonselfaveraging 
of the overlap. Of course, the system possesses spin-flip symmetry
also for $\lambda\neq0$, so that $\med{q_{12}}=0$. Therefore,
if $\med{q_{12}^2}\neq0$ then the overlap is nonselfaveraging. Now
replica symmetry is explicitly broken. 

The basic idea of our method is to show that, as long as $t<1$ and $\lambda$ is
small enough, the term $\lambda N q_{12}^2$  does not change the value of the
free energy in the thermodynamic limit. Therefore, ``most'' configuration must
have $q_{12}=0$ and the overlap  must be selfaveraging.
In order to implement this intuitive idea, one introduces the $\lambda$ 
dependent auxiliary function
$$\tilde\alpha_N(t,\lambda)=\frac1{2N}E\ln\tilde Z_N,$$
where the normalization factor $1/2$ is chosen so that
$\tilde \alpha_N(t,0)=\alpha_N(t)$.
Through a simple explicit calculation, we can easily calculate the $t$
derivative in the form
\begin{equation}
\label{tder}
\partial_t\tilde\alpha_N=\frac14\left(
1+\med{q_{12}^2}-2\med{q_{13}^2}
\right),
\end{equation}
where now all averages $\med{\,}$ involve the $\lambda$-dependent state with {\it
Boltzmannfaktor} given in agreement with (\ref{tildez}).
Moreover, it is obvious that 
$$\partial_{\lambda}\tilde\alpha_N=\frac14\med{q_{12}^2}.$$
%
Starting from some point $\lambda_0>0$, consider the linear trajectory
$\lambda(t)=\lambda_0-t$, with obvious invertibility in the form 
$\lambda_0=\lambda+t$. Let us take the $t$ derivative of $\tilde\alpha_N$
along this trajectory
%
$$\frac d{dt}\tilde\alpha_N(t,\lambda(t))=\left(
\partial_t-\partial_{\lambda}\right)\tilde\alpha_N=\frac14-\frac12
\med{q_{13}^2}_{t,\lambda(t)}.$$
%
Notice that the term containing $\med{q_{12}^2}$ disappeared. By integration
we get the sum rule and the inequality
\begin{eqnarray}
\label{sumru}
\tilde\alpha_N(t,\lambda)=\frac t4+\tilde\alpha_N(0,\lambda_0)
-\frac12\int_0^t\med{q_{13}^2}_{t',\lambda(t')}\,dt'\leq 
\frac t4+\tilde\alpha_N(0,\lambda_0),
\end{eqnarray}
where $\med{q_{13}^2}_{t',\lambda(t')}$ refers to 
$\lambda(t')=\lambda_0-t'=\lambda+t-t'$.

Next, we compute $\tilde\alpha_N(0,\lambda_0)$. We introduce an
auxiliary unit Gaussian $z$, and perform simple rescaling, in order to obtain 
\begin{eqnarray}
\label{gauss}
\tilde\alpha_N(0,\lambda_0)&=&\frac1{2N}\ln\sum_{\{\sigma,\sigma'\}}
e^{\frac12\lambda_0Nq_{12}^2}=\frac1{2N}\ln\sum_{\{\sigma,\sigma'\}}
\int e^{\sqrt {\lambda_0 N}q_{12}z}\,d\mu(z)\\\nonumber
&=&\ln2+\frac1{2N}\ln\int \left(\cosh\,z\sqrt{\frac{\lambda_0}N}\right)^N
\, d\mu(z)\\
\label{sella}
&=&\ln2+\frac1{2N}\ln
\int dy\sqrt{\frac{N \lambda_0}{2\pi}} \exp N\left(-\lambda_0\frac{y^2}2+
\ln\cosh(y\lambda_0)\right),
\end{eqnarray}
where we performed the change of variables $z$ to $y\sqrt{N \lambda_0}$ in 
the last step. It is immediately recognized that the integral in (\ref{sella})
appears in the ordinary treatment of the well known ferromagnetic mean field
Curie-Weiss model. The saddle point method gives immediately
\begin{equation}
\label{azero}
\lim_{N\to\infty}\tilde\alpha_N(0,\lambda_0)=\ln2+\frac12 
\max_y \left(-\lambda_0\frac{y^2}2+
\ln\cosh(y\lambda_0)\right).
\end{equation}
Therefore, the critical value for $\lambda_0$ is $\lambda_c=1$. For
$\lambda_0>1$ we have 
$$\lim_{N\to\infty}\tilde\alpha_N(0,\lambda_0)>\ln2,$$
%
while for $\lambda_0<1$, one
can use the elementary property $2\ln\cosh x\leq x^2$ to find
\begin{eqnarray}
\label{sottosoglia}
\tilde\alpha_N(0,\lambda_0)\leq\ln2+\frac{1}{4N}\ln\frac{1}{1-\lambda_0}.
\end{eqnarray}
Notice that, when $\lambda_0$ approaches the value $1^-$, the term of
order $1/N$ diverges, since Gaussian fluctuations around the saddle point 
become larger and larger. 

Thanks to (\ref{sottosoglia}), the inequality in (\ref{sumru}) becomes
%
$$\tilde\alpha_N(t,\lambda)\leq{\overline\alpha}(t)+
\frac{1}{4N}\ln\frac{1}{1-\lambda_0},$$
%
which holds for $0\leq\lambda_0<1$, i.e., for $0\leq t+\lambda<1$.

Next, we use convexity of $\tilde\alpha_N(t,\lambda)$ with respect 
to $\lambda$ and the fact that 
$$\left.\partial_\lambda\tilde\alpha_N(t,\lambda)\right|_{\lambda=0}=
\frac14\med{q_{12}^2}_t$$
to write
$$\alpha_N(t)+\frac\lambda4\med{q_{12}^2}_t\leq\tilde\alpha_N(t,\lambda)
\leq{\overline\alpha}(t)+\frac{1}{4N}\ln\frac{1}{1-\lambda-t},$$
for $\lambda>0$.
For $0\leq t\leq \bar t<1$, choose $\lambda=(1-\bar t)/2$, so that
$$\lambda+t\leq \bar \lambda_0\equiv(1+\bar t)/2<1,$$
and
\begin{equation}
\label{accorc}
\frac14\med{q_{12}^2}_t\leq\frac1\lambda({\overline\alpha}(t)-\alpha_N(t))
+\frac{1}{4N\lambda}\ln\frac{1}{1-\bar\lambda_0}.
\end{equation}
Recalling Eq. (\ref{rs}), one has
\begin{equation}
\frac d{dt}({\overline\alpha}(t)-\alpha_N(t))=\frac14\med{q_{12}^2}_t\leq
\frac1\lambda({\overline\alpha}(t)-\alpha_N(t))+
\frac{1}{4N\lambda}\ln\frac{1}{1-\bar\lambda_0},
\end{equation}
so that
$$\alpha_N(t)={\overline\alpha}(t)+O(1/N),$$
uniformly for $0\leq t\leq\bar t<1$.
Of course, from Eq. (\ref{sumru}) and convexity of $\tilde\alpha_N$ one 
also has
\begin{eqnarray}\nonumber
&&\tilde\alpha_N(t,\lambda)={\overline\alpha}(t)+O(1/N),\\\nonumber
&&\med{q_{13}^2}_{t,\lambda}=O(1/N),
\end{eqnarray}
for $0\leq t+\lambda\leq\bar\lambda_0<1$.

We have gained a complete control of the system in the triangular region
$0\leq t<1$, $0\leq \lambda<1-t$.
Note that we have not only proved Eq. (\ref{ann}) but we also shown that
the leading correction to annealing is of order at most $1/N$.

\section{The general case}

The method we follow for the general case, where the {\it Boltzmannfaktor} is
given by (\ref{Bf}), is a direct generalization of
the one explained in the previous section. In fact, by taking into account the
$t$ derivative in  (\ref{veri}), we are led to
introduce the auxiliary function  
$$\tilde\alpha_N(x,\lambda,t)=\frac1{2N}{E}\ln \tilde Z_N(x,\lambda,t;J),$$
where $\tilde Z_N$ is the partition function for a system of two replicas 
coupled by the term
$$\frac\lambda2N(q_{12}-\bar q(x,t))^2,$$
with $\lambda\ge0$.
In order to simplify notation, we omit the argument $h$.

Now the $t$ derivative is given by
\begin{eqnarray}\nonumber
&&\partial_t\tilde\alpha_N=\frac14\left(1+\med{q_{12}}-2\med{q_{13}}\right)
+\frac\lambda2(\bar q-\med{q_{12}})\frac{\partial\bar q}{\partial t},
\end{eqnarray}
while the $x$ and $\lambda$ derivatives appear as
\begin{eqnarray}\nonumber
&&\partial_x\tilde\alpha_N=\frac12\left(
1+\med{q_{12}}-2\med{q_{13}}\right)+
\frac\lambda2(\bar q-\med{q_{12}})\frac{\partial\bar q}{\partial x},
\\\nonumber
&&\partial_\lambda\tilde\alpha_N=\frac14\Med{(q_{12}-\bar q(x,t))^2}.
\end{eqnarray}
Starting from points $\lambda_0>0$, $x_0$, consider the linear trajectories
$\lambda(t)=\lambda_0-t$, $x(t)=x_0-\bar q(x_0)t$, as in (\ref{traj0}), with
obvious invertibility as explained before. Since $\bar q$ is constant 
along the trajectory, one finds
for the total time derivative of $\tilde\alpha_N$ 
$$\frac d{dt}\tilde\alpha_N(x(t),\lambda(t),t)=\left(
\partial_t-\bar q\,\partial x
-\partial_{\lambda}\right)\tilde\alpha_N=\frac14(1-\bar q)^2
-\frac12\Med{(q_{13}-\bar q)^2}.$$
Notice that in this case the term containing
$\Med{(q_{12}-\bar q)^2}$ disappeared. 

By integration, we get the sum rule and
the inequality 
\begin{eqnarray}
\nonumber 
\tilde\alpha_N(x,\lambda,t)&=&\frac
t4(1-\bar q)^2+\tilde\alpha_N(x_0,\lambda_0,0)
-\frac12\int_0^t\Med{(q_{13}-\bar q)^2}_{t',x(t'),\lambda(t')}\ dt'\\\nonumber
&\leq & \frac
t4(1-\bar q)^2+\tilde\alpha_N(x_0,\lambda_0,0),
\end{eqnarray}
where $\Med{(q_{13}-\bar q)^2}_{t',x(t'),\lambda(t')}$ refers to 
\begin{eqnarray}\nonumber
\lambda(t')&=&\lambda_0-t'=\lambda+t-t',\\\nonumber
x(t')&=&x_0-\bar q t'=x+\bar q(t-t') .
\end{eqnarray}
If $\Omega_J$ is the product state for two replicas with the original
{\it Boltzmannfaktor} given by (\ref{Bf}), then we can write
$$\tilde\alpha_N(x,\lambda,t)-\alpha_N(x,t)\equiv 
\frac1{2N}E \ln \Omega_J \left(\exp {\frac12\lambda N (q_{12}-\bar q)^2}\right)
.$$
Therefore, by exploiting Jensen inequality, we have, for $\lambda\ge0$,
$$\frac{\lambda}{4}\Med{(q_{12}-\bar q)^2}_{x,t}\le
\tilde\alpha_N(x,\lambda,t)-\alpha_N(x,t).$$

Let us also define
\begin{equation}
\label{Delta}
\Delta_N(x_0,\lambda_0)\equiv
\tilde\alpha_N(x_0,\lambda_0,0)-\alpha(x_0,0)=
\frac1{2N}E \ln \Omega_J^0 \left(\exp {\frac12\lambda_0 N (q_{12}-\bar
q)^2}\right),
\end{equation}
where we have introduced the state $\Omega_J^0$ for two replicas,
corresponding to $t=0$, and $x=x_0$, in (\ref{Bf}). Notice that
$\Omega_J^0$ is a factor state over the sites $i$.

By collecting all our definitions and inequalities, and taking into account
the definition (\ref{aSK}), we have
$$\frac{\lambda}{4}\Med{(q_{12}-\bar q)^2}_{x,t}\le
\Delta_N(x_0,\lambda_0)+\bar\alpha(x,t)-\alpha_N(x,t).$$

Let us now introduce $\lambda_c(x_0)$ such that, for any 
$\lambda_0\le\lambda_c(x_0)$, one has
$$\lim_{N\to\infty}\Delta_N(x_0,\lambda_0)=0.$$
Then, by the same reasoning already exploited starting from (\ref{accorc}),
and taking into account (\ref{veri}), we obtain the proof of the following
\begin{theorem}
For any $t\le\lambda_c(x_0(x,t))$, where $x_0(x,t)$ is defined as in Theorem
\ref{unique}, we have the convergence \begin{equation}
\lim_{N\to\infty}\alpha_N(x,t)=\bar\alpha(x,t).
\end{equation}
\end{theorem}

For the specification of $\lambda_c(x_0)$, we can easily establish the complete
characterization of the $\Delta_N$ limit. In fact, the following
holds.
\begin{theorem} 
\label{DeltaN}
The infinite volume limit of $\Delta_N$ is given by  
$$\lim_{N\to\infty}\Delta_N(x_0,\lambda_0)=\Delta(x_0,\lambda_0).$$
Here, $\Delta(x_0,\lambda_0)$ is defined through the variational expression
\begin{equation}
\label{max}
\Delta(x_0,\lambda_0)\equiv
\frac12\max_\mu\left(\int\ln(\cosh\mu+\tanh^2(\beta h + z \sqrt{x_0})\sinh\mu)
d\rho(z) -\mu\bar q-\frac{\mu^2}{2\lambda_0}\right),
\end{equation}
where $d\rho(z)$ is the centered unit Gaussian measure.
\end{theorem}

Of course, the expression (\ref{max}) is in agreement with (\ref{azero}), when there are no external fields.

It is easy to realize that  the value $\lambda_c$, in the general case, is strictly
less than the expected value $t_c$, following  from  the Almeida-Thouless argument,
$$t_c\int\cosh^{-4}(\beta
h + z \sqrt{x_0}) d\mu(z)=1.$$

The proof of the Theorem \ref{DeltaN} is easy. First of all let us establish 
the elementary bound, uniform in $N$,
\begin{equation}
\label{bound}
\Delta_N(x_0,\lambda_0)\ge \Delta(x_0,\lambda_0).
\end{equation}
In fact, starting from the definition of $\Delta_N(x_0,\lambda_0)$ given in (\ref{Delta}), we can write, for $\lambda_0\ne0$, and any $\mu$,
$$(q_{12}-\bar q)^2\ge 2\frac{\mu}{\lambda_0}(q_{12}-\bar q)-\left(\frac{\mu}{\lambda_0}\right)^2,$$
and conclude that
\begin{equation}
\label{boundmu}
\Delta_N(x_0,\lambda_0)\ge \alpha_0(\mu) - \frac{\mu^2}{4\lambda_0},
\end{equation}
where we have defined
\begin{eqnarray}
\nonumber
\alpha_0(\mu)&\equiv& \frac1{2N}E \ln \Omega_J^0 \left(\exp {\mu N (q_{12}-\bar
q)}\right)\\\nonumber
&=&\frac12 \int \ln \left(\cosh\mu+ \tanh^2(\beta
h + z \sqrt{x_0})\sinh\mu\right)\ d\rho(z)-\frac12 \mu\bar q. 
\end{eqnarray}
Of course, it is convenient to take the $\max_\mu$ in the r.h.s. of (\ref{boundmu}), so that the bound in (\ref{bound}) is established. The proof that the bound is in effect the limit, as $N\to\infty$, can be obtained in a very simple way by using a Gaussian transformation on (\ref{Delta}), as it was done in (\ref{gauss}). In fact, we now have
\begin{eqnarray}
&&\frac1{2N}E \ln \Omega_J^0 \left(\exp {\frac12\lambda_0 N (q_{12}-\bar
q)^2}\right)\\\nonumber
&&=\frac1{2N}E \ln \int \Omega_J^0 \left(\exp {\sqrt{\lambda_0 N} (q_{12}-\bar
q) z}\right)\ d\rho(z)
\end{eqnarray}
Therefore, by exploiting the fact that also $\Omega_J^0$ factorizes with respect to the sites $i$, we can write
\begin{eqnarray}
\label{dmuz}
\nonumber
\Delta_N(x_0,\lambda_0)\!\!&=&\!\!\frac1{2N}E\ln\int\prod_i
\left(\cosh\sqrt{\frac{\lambda_0}{N}}z+\tanh^2(\beta h + J_i\sqrt{x_0})
\sinh\sqrt{\frac{\lambda_0}{N}}z\right) \\
&&\times\exp\left(-\sqrt{\lambda_0 N}\bar q z\right)\ d\rho(z).
\end{eqnarray} 
Now, we find convenient to introduce a small $\epsilon>0$, so that
\begin{equation}
\label{lambdaprimo}
\frac1{\lambda_0}=\frac1{\lambda_0^{\prime}}+\epsilon.
\end{equation}
Notice that $\lambda_0<\lambda_0^{\prime}$.
We also introduce the auxiliary (random) function
$$\phi_N(y,\lambda_0^{\prime})\equiv \frac1N\sum_i\ln
\left(\cosh y+\tanh^2(\beta h + J_i\sqrt{x_0})\sinh y\right)-\bar q y -\frac12\frac{y^2}{\lambda_0^{\prime}}.$$
By the strong law of large numbers, as $N\to\infty$, for any $y$, we have the $J$ almost sure convergence of $\phi_N(y,\lambda_0^{\prime})$ to $\phi(y,\lambda_0^{\prime})$ defined by
\begin{eqnarray}
\nonumber
\phi(y,\lambda_0^{\prime})&\equiv& \int \ln
\left(\cosh y+\tanh^2(\beta h + z\sqrt{x_0})\sinh y\right) d\rho(z)-
\bar q y -\frac12\frac{y^2}{\lambda_0^{\prime}}\\\nonumber
&=&
E\phi_N(y,\lambda_0^{\prime}).
\end{eqnarray}
Here $d\rho(z)$ performs the averages with respect to the $J_i$ variables.
Let us also remark that the convergence is $J$ almost surely uniform for any finite number of values of the variable $y$.

Now we can go back to (\ref{dmuz}), write explicitly the unit Gaussian measure $d\rho(z)$, perform the change of variables $y=z\sqrt{\lambda_0 N^{-1}}$, make the transformation (\ref{lambdaprimo}), take the $\sup_y$ for the $\phi_N$, and perform the residual Gaussian integration over $y$. We end up with the estimate
\begin{equation}
\label{esup}
\Delta_N(x_0,\lambda_0)\le \frac12 E \sup_y \phi_N(y,\lambda_0^{\prime})+\frac1{2N}\ln \frac1{\sqrt{\lambda_0 \epsilon}}.
\end{equation}
Since the $J$-dependent $\sup_y$ is reached in some finite interval, for any fixed $\lambda_0^{\prime}$, and the function $\phi_N$ is continuous with respect to $y$, with bounded derivatives, we can perform the $\sup_y$ with $y$ running over a finite discrete mesh of values, by tolerating a small error, which becomes smaller and smaller as the mesh interval is made smaller. But in this case the strong law of large numbers allows us to substitute $\phi_N$ with $\phi$, in the infinite volume limit $N\to\infty$. On the other hand, the second term in the r.h.s. of (\ref{esup}) vanishes in the limit. Therefore, we conclude that
$$\limsup_{N\to\infty}\Delta_N(x_0,\lambda_0)\le \frac12\sup_y \phi(y,\lambda_
0^{\prime}).$$
From continuity with respect to $\lambda_0$, we can let $\lambda_0^{\prime}$ approach $\lambda_0$, and the Theorem is proven.

\section{Fluctuations of overlaps and free energy}

In the previous sections we proved that, in a certain region of the 
parameters $t,x,\beta h$, the typical values of the free energy $\ln Z_N/N$ and
of the overlap $q_{ab}$ are  the replica symmetric expressions $\bar\alpha$ 
and $\bar q$, respectively. In the same region, one can obtain a more precise
characterization of the fluctuations of these quantities, for 
$N\to\infty$, showing that a central limit-type
theorem holds, after suitable rescaling.
This will be analyzed in detail in a subsequent
paper \cite{GT}. Here, we just give the main results and sketch 
the ideas underlying the proof.

Concerning the fluctuations of the overlap around the Sherrington-Kirkpatrick
order parameter $\bar q$, we prove the following
\begin{theorem}\cite{GT}
\label{teofluthQ}
The rescaled overlap variables
$$\xi^N_{ab}\equiv \sqrt N (q_{ab}-\bar q)$$
tend in distribution, as $N\to\infty$, to centered, jointly Gaussian 
variables with covariances 
\begin{eqnarray}
\nonumber
&&\med{\xi^2_{ab}}=A(t,x,h)\\\nonumber
&&\med{\xi_{ab}\xi_{ac}}=B(t,x,h)\\\nonumber
&&\med{\xi_{ab}\xi_{cd}}=C(t,x,h),
\end{eqnarray}
where $b\neq c$, $c\neq a,b$ and $d\neq a,b$. The expressions of 
$A,B,C$ are explicitly given and coincide with those found 
in \cite{SUM}.
\end{theorem}
Recently, an analogous 
result was proved independently by Talagrand \cite{talalibro}, who
computed the $N\to\infty$ limit for all moments of the $\xi$ variables. 

The scheme of our proof is as follows: The control we obtained
on the coupled two replica system and concentration of measure inequalities for
the free energy \cite{talagcorso} imply that the 
fluctuations of 
$q_{ab}$ from $\bar q$ are exponentially suppressed for $N$ large. Then, by 
means of the
cavity method \cite{MPV} one can write a self-consistent closed equation
for the characteristic function of the variables $\xi^N_{ab}$. This equation
turns out to be linear, apart from
error terms which vanish asymptotically for $N\to\infty$, thanks to the 
strong suppression of the overlap fluctuations. The solution, which is
easily found, coincides with the characteristic function of a 
Gaussian distribution, with the correct covariance structure. 

Concerning the free energy, the result we prove is the following:
\begin{theorem}\cite{GT}
\label{fluttZ}
Define the rescaled free energy fluctuation as
$$\hat f_N(t,x,h;J)\equiv\sqrt N\left(\frac{\ln Z_N(t,x,h;J)}N-
\bar\alpha(t,x,h)\right).$$
Then,
$$\hat f_N(t,x,h;J)\stackrel{d}{\longrightarrow} {\mathcal N}(0,\sigma^2
(t,x,h)),
$$
where ${\cal N}(m,\sigma^2)$ denotes the Gaussian random variable of mean 
$m$ and variance $\sigma^2$, and
$$\sigma^2(t,x,h)={\mbox Var}\left(\ln\cosh(z\sqrt{\bar q t+x}+\beta h)\right)-
\frac{\bar q ^2 t}2.$$
Here, $Var(.)$ denotes the variance of a random variable and $z={\mathcal N}
(0,1)$.
\end{theorem}
This result is a consequence of Theorem \ref{teofluthQ} and of concentration 
of measure inequalities for the free energy.

\section{Outlook and conclusions}

We obtained control on the thermodynamic limit of the model, in a region
above the Almeida-Thouless line, by  suitably coupling two replicas of the 
system and studying stability with respect to the coupling parameter. 
The question naturally arises, whether and how this method can be 
extended up to the expected critical line. This problem seems to be
common to all approaches proposed so far.

The method can be also further generalized to the case where more and more
replicas are mutually coupled. In this case, 
replica symmetry is explicitly broken at various levels, and it is possible to 
give a generalization of the Ghirlanda-Guerra relations \cite{GG}. 
We plan to report soon on these generalizations \cite{ancoranoi}.

\vspace{.5cm}
{\bf Acknowledgments}

We gratefully acknowledge useful conversations with
Pierluigi Contucci, Roberto D'Autilia, Sandro Graffi, Enzo Marinari, Giorgio
Parisi, Masha Scherbina, and Michel Talagrand.

This work was supported in part by MIUR 
(Italian Minister of Instruction, University and Research), 
and by INFN (Italian National Institute for Nuclear Physics).

\end{document}